\documentclass[aps,apl,twocolumn,showpacs,amsmath,floatfix,superscriptaddress]{revtex4}

\usepackage{graphicx}

\begin{document}

\title{Optical Absorption Spectra of Electrically Gated Bilayer Graphene}

\author{Li Yang}
\affiliation{Department of Physics, Washington University, St.
Louis, Missouri, 63130, USA}

\date{\today}

\begin{abstract}
The electronic structure and optical response of electrically
gated bilayer graphene are studied by first-principles approaches.
We have obtained the induced band gap that is in good agreement
with experiment when the applied electric field is less than 1.5
V/nm. The infrared optical absorbance is calculated within the
single-particle excitation picture and its fine structures are
presented. In addition, the calculated infrared optical absorbance
is found to be strongly depending on stacking styles of bilayer
graphene and the polarization direction of the incident light,
which provides efficient ways to identify the electric-field
intensity and stacking styles in experiment. Finally,
many-electron effects are discussed.
\end{abstract}


\maketitle


\section{Introduction}

Graphene, a single-layer graphite, is known for its unique
electronic structure that has a massless Dirac-fermion dispersion
close to the Fermi level \cite{geim-1,geim-2,kim05,heer}. This
special feature results in many unusual properties
\cite{rev-1,rev-2}, e.g., quantum-Hall effect \cite{kim05}, Kohn
anomaly \cite{kohn-1,kohn-2} and universal infrared optical
conductance
\cite{ando02,sharapov06,neto06,nair08,heintz08,dawlaty08,
wright09}, etc.. In addition to the importance of fundamental
physics, the high mobility of free carriers and 2-dimensional
(2-D) nature of graphene make it possible to obtain
high-performance microelectronic circuit structures, which could
dramatically simplify the fabrication of devices and lower the
cost consequently. However, despite above outstanding properties,
one obstacle to applications of graphene is its zero-gap band
structure. As a result, electrical conduction cannot be turned off
using control voltages, which is essential for the operation of
transistors \cite{geim06}.

Recent experiments have confirmed that an external electric field
perpendicularly applied to bilayer graphene (BLG) can modify the
electronic structure and induce a finite band gap by breaking the
lattice inversion symmetry without degrading the high mobility
\cite{ohta06,chen08,oos08}. Moreover, the induced band gap can be
efficiently tuned in a wide range, up to a few tenths of an eV by
the applied field around 1$\sim$2 V/nm \cite{wang09,mak09}. This
discovery makes BLG the first known material with a wide-range
tunable band gap. On the other hand, many theoretical studies have
been performed to reveal the band structure of electrically gated
BLG \cite{neto07,min07,lin06,nicol08,mauri09,mo09}. However, there
are very few first-principles calculations about its optical
absorption spectrum although a number of relevant experiments are
using optical approaches to study BLG \cite{wang09,mak09} and
corresponding tight-binding (TB) models \cite{tb-2} have been
developed. Therefore, a first-principles calculation about the
optical response of BLG is of great interest to the graphene
community.

Beyond explaining available experimental data, we are motivated to
study polarization effects of the optical response of electrically
gated BLG because low-dimensional materials usually display quite
different optical response to the incident light with different
polarization directions
\cite{ando94,catalin04,eric04,yang07,zhang08}. Therefore, the
optical absorption spectra by different polarization directions
may provide useful information to detect atomic and electronic
structures of BLG \cite{ando09}. Unfortunately, to date we have
very limited first-principles knowledge about the polarization
dependence of the infrared optical absorbance of electrically
gated BLG. Finally, the stacking style of BLG is another
interesting topic affecting its infrared optical response because
different stacking styles result in different band structures
around the Dirac point \cite{gonze1}. In particular, experimental
conditions, such as external strain, imperfections and edges, can
potentially induce different stacking styles locally
\cite{edge-1}. Since it is not easy to identify the stacking style
of BLG in experiment, the optical measurement may provide an
efficient way to solve this problem.

Motivated by above considerations, we have performed
first-principles calculations to study the electronic structure
and optical absorption spectra of AA and AB stacked BLG with an
perpendicularly applied electric field. Our DFT-calculated band
gap is in good agreement with experimental measurements when the
applied electric field is weak. Correspondingly, this induced band
gap results in a significant modification of optical absorption
spectra of BLG within the infrared frequency regime. Our
calculation reveals that not only absorption peak positions but
also their amplitudes and lineshapes are significantly changed by
the applied field. In addition, electrically gated BLG displays
strong polarization effects and a dependence of stacking styles,
which are useful for its future electronic and photonic
applications.

The remainder of this paper is organized as follows: in section
II, we introduce the calculation details and structure of
electrically gated BLG; in section III, the band gap of
electrically gated BLG and comparisons with experimental
measurements are presented; in section IV, we carry out detailed
calculations on optical absorption spectra of electrically gated
BLG with different stacking styles and polarization directions; in
section V, many-electrons effects on infrared optical absorption
spectra are discussed; in section VI, we summarize our studies and
conclusion.

\section{Structure and Calculation Details}

Our calculations are using density functional theory (DFT) within
the local density approximation (LDA) \cite{kohn-a,kohn-b} and the
computational package is Quantum ESPRESSO \cite{pwscf}. The
calculations are done in a supercell arrangement \cite{cohen75}
with a plane-wave basis using norm-conserving pseudopotentials
\cite{martin91} with an 80 Ry energy cutoff. The distance between
BLG sheets in neighboring supercells is set to be 2.0 nm to avoid
spurious interactions. Two valence bands and two conduction bands
are included to obtain converged optical absorption spectra up to
6 eV. A saw-tooth shape of electric potential is perpendicularly
applied to mimic the gating electric field. A 128x128x1 k-point
grid is used to ensure converged DFT results. In this work, we
focus on isolated BLG with fixed chemical potential, although the
applied field can also be used to modify the chemical potential
and induces a substantial change of optical absorption spectra
\cite{wang08}. Thus the optical response is studied by calculating
the imaginary part of the dielectric function \cite{cohenbook}
\begin{equation}\label{1}
\varepsilon_2(\omega) = \frac{16\pi e^2}{\omega^2}\sum_{v,c} |\vec
\lambda \cdot \langle v | \vec v | c \rangle |^2 \delta (\omega -
(E_c-E_v)),
\end{equation}
where $|v\rangle$ and $|c\rangle$ are valence and conduction
states, respectively, $\vec v$ is the velocity operator and $\vec
\lambda$ is the polarization direction of the incident light.

However, the quantity of above imaginary dielectric function
cannot be compared with experiments directly because its value is
depending on the choice of the supercell size. In order to
eliminate this artificial effect, we obtain the polarizability per
unit area of BLG by \cite{ch06}
\begin{equation}\label{2}
\alpha_2(\omega)=(\varepsilon(\omega)-1)d/4\pi,
\end{equation}
where $d$ is the distance between adjacent BLG sheets in our
supercell arrangement. Moreover, most of experimental measured
quantities are the optical absorbance. If we assume isolated BLG
surrounded by infinite vacuum, the optical absorbance can be
derived as \cite{yang09}
\begin{equation}\label{3}
A(\omega)=\frac{4\pi\omega}{c}\alpha_2(\omega)= \frac{16\pi e^2
d}{\omega c}\sum_{v,c} |\vec \lambda \dot \langle v | \vec v | c
\rangle |^2 \delta (\omega - (E_c-E_v)).
\end{equation}

The other challenge of this study is how to obtain an optical
absorbance with a good energy resolution. Because available
experiments can only induce a small band gap of BLG in an order of
a few tenths of an eV \cite{oos08,wang09,mak09}, we have to use an
extremely dense k-point sampling to obtain comparable optical
absorption spectra. Fortunately, we are interested in the infrared
optical absorption spectrum up to 1 eV and only need to
extensively sample the k-space around the Dirac point. In this
study, we use a 100x100 k-point grid to sampling the mini first
Brillouin zone (BZ) (0.1x0.1 of the first BZ) around the Dirac
point, which is equivalently a 1000x1000 k-point sampling of the
whole first BZ. This extremely dense sampling makes it possible to
obtain a fine structure of optical absorption spectra with a 20
meV energy resolution.

Finally, we have considered two stacking styles of BLG, AB and AA.
All these electrically gated structures are fully relaxed
according to the atomic force and stress with DFT/LDA. We find
that the applied electric field has minor effects on the structure
of BLG. The relaxed inter-layer distance and C-C bond length are
nearly identical under different electric fields up to 4 V/nm. The
relaxed inter-layer distance of AB stacked BLG is 0.335 nm and
that of AA stacked one is 0.346 nm, respectively, which are
consistent with previous first-principles results \cite{penn-1}.

\section{Band Gaps of Electrically Gated BLG}

\begin{figure}
\includegraphics*[scale=0.85]{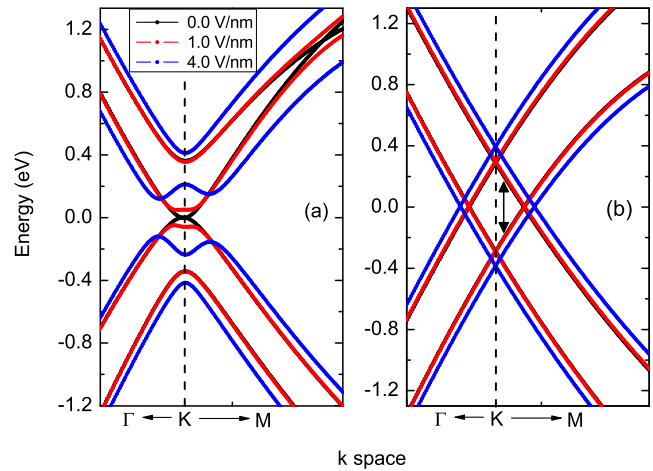}
\caption{\label{fig:band-2} (Color online) Band structure of
electrically gated BLG close to the Dirac point. The AB stacked
one is shown in (a) and AA stacked on is shown in (b).}
\end{figure}

Plotted in Fig.~\ref{fig:band-2} is the band structure of BLG
close to the Dirac point calculated by DFT/LDA. In the absence of
gating field, for AB stacked one shown in Fig.~\ref{fig:band-2}
(a), the valence band and conduction band touch each other with a
quadratic shape due to breaking the AB symmetry by inter-layer
interactions. Moreover, dispersions of valence bands and
conduction bands are not symmetric to each other according to the
Dirac point \cite{li09}. In particular, the lowest two conduction
bands are even crossing each other along the K-M direction. This
may induce impacts on optical absorption spectra of doped BLG
\cite{wang08}. For AA stacked BLG shown in Fig.~\ref{fig:band-2}
(b), the inter-layer interaction does not change the band
dispersion but shifts Dirac points.

When gating electric field is applied, a finite band gap is
generated in AB stacked BLG. It can be understood by the following
Hamiltonian describing the electronic structure near the Dirac
point of AB stacked BLG \cite{tb-1},
\begin{equation}\label{2}
H=\left(%
\begin{array}{cc}
  \Delta & -\frac{\hbar^2}{2m}(k_x-ik_y)^2 \\
  -\frac{\hbar^2}{2m}(k_x+ik_y)^2 & -\Delta \\
\end{array}
\right)
\end{equation}
where $k$ is the momentum and $\Delta$ is the onsite energy
difference between two layers of BLG, respectively. In the absence
of electric field, $\Delta=0$, thus the above effective
Hamiltonian will lead to a gap-less quadratic band dispersion.
When gating electric field is applied, it will introduce different
onsite energies of two layers. Then the non-zero $\Delta$ will
give rise to a finite band gap with a size of 2$\Delta$. However,
how to obtain the value of $\Delta$ is not easy because the
applied electric field is inevitable to be screened by electrons
in graphene, which can significantly depress the difference of
onsite energy between graphene layers and reduce the band gap.
DFT/LDA may be a better choice because it includes a part of
screening effects through first-principles ways.

In Fig.~\ref{fig:band-2} (a), as the intensity of the applied
electric field increases, the band gap is enlarged and the band
structure is no longer quadratic and finally replaced by a
Mexican-hat shape dispersion. In addition to the induced band gap,
the dispersion of conduction bands is modified by the applied
field as well. For example, the lowest two conduction bands are no
longer crossing each other along the K-M direction under strong
applied field. On the contrary, the bottom of the second lowest
conduction band and the top of the second highest valence band are
not so sensitive to the applied electric field, and a significant
change of these bands shows up until the applied field is larger
than 4 V/nm.

\begin{figure}
\includegraphics*[scale=0.95]{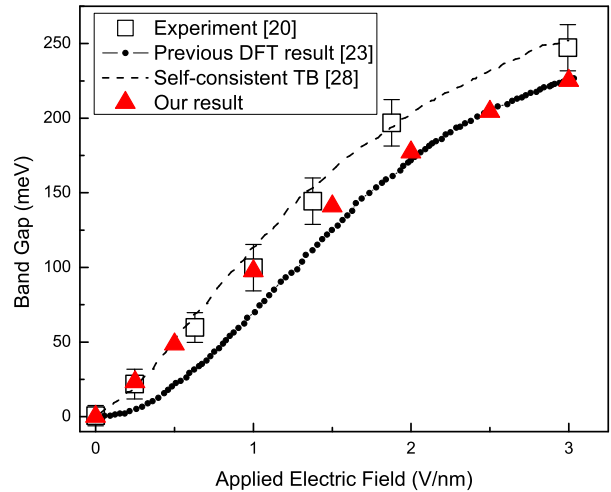}
\caption{\label{fig:compare-1} (Color online) Electric-field
dependence of the tunable band gap in AB stacked BLG. The
experimental measurements and tight-binding and previous \emph{ab
initio} results are retrieved from Fig. 4 of Ref. \cite{wang09}.}
\end{figure}

We summarize our calculated band gap under different electric
fields into Fig.~\ref{fig:compare-1}. Previous self-consistent TB
\cite{tb-2} and \emph{ab initio} calculations \cite{min07} and
experimental measurements \cite{wang09} are plotted together for
comparison. Interestingly, our calculated band gap is larger than
previous \emph{ab initio} calculations \cite{min07} and in good
agreement with experimental measurements when gating field is less
than 1.5 V/nm. For example, the previous DFT calculated band gap
under a 0.5 V/nm field is around 30 meV, but our calculation
provides a 50 meV gap, an over 60\% enlargement. A larger energy
cutoff and denser k-grid used in our calculations may be reasons
for the difference between ours and the previous calculation
\cite{min07}. We have checked some other first-principles studies
and find that they are consistent with our results
\cite{mauri09,mo09}. For example, following Ref. \cite{mauri09},
the band gap of BLG is 37 meV when the applied field is 0.45 V/nm,
which is very close to our result. In Ref. \cite{mo09}, they get a
band gap around 100 meV when the applied electric field is 1 V/nm,
which is consistent with our data as well. For larger electric
field ($>$ 1.5 V/nm), our calculated band gap is approaching
previous \emph{ab initio} results \cite{min07} and significantly
smaller than experimental measurement.

The reason for the consistence and inconsistence of our calculated
band gap with experimental data is complicated because DFT is
known for its failure to obtain accurate band gaps of
semiconductors \cite{gw}. In particular, there are studies
supporting that self-energy corrections may enlarge the band gap
of BLG \cite{yang09,mauri09}. Here we attribute this ``partial
success" of DFT to experimental reasons, such as substrate
effects. In experiment, BLG is sandwiched between gates, which may
significantly enhance the screening between electrons. Thus this
factor can reduce self-energy corrections from many-electron
effects and makes DFT results close to experimental measurements.
However, when the band gap is large enough as the applied field is
more than 1.5 V/nm, substrate effects come to be smaller than
self-energy corrections and our DFT result starts to significantly
underestimate the band gap as shown in Fig.~\ref{fig:compare-1}.
Another potential reason for the above agreement between DFT gaps
and optical gaps may be from the cancellation between self-energy
corrections and excitonic effects \cite{yang07}. More accurate
experiments and first-principles calculations with many-electron
effects included are expected to verify our discussions.

The band structure of AA stacked BLG close to the Dirac point is
presented in Fig.~\ref{fig:band-2} (b). Because inversion and AB
symmetries are kept, we do not observe any finite band gap even
when the applied electric field is around 4 V/nm. However, we do
see an enlargement of the separation between two Dirac points
marked in Fig.~\ref{fig:band-2} (b), which is a result from the
enhancement of the difference of onsite energy of two layers due
to gating field. This small but essential change of band structure
will result in corresponding modifications of optical absorption
spectra and will be discussed in the next section.

\section{Optical Absorption Spectra of Electrically Gated BLG}

\begin{figure}
\includegraphics*[scale=0.85]{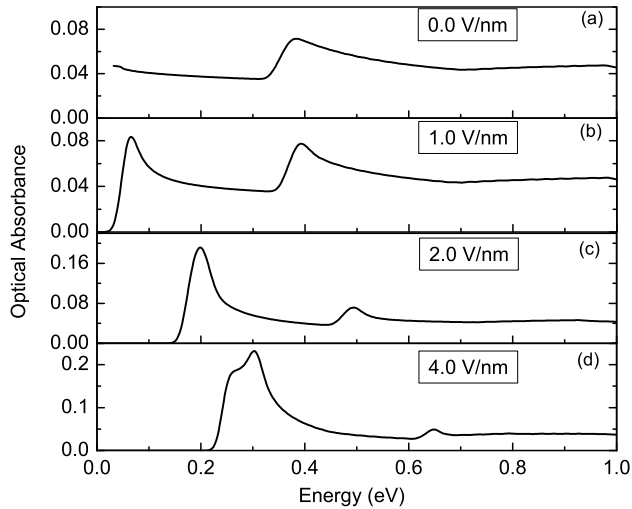}
\caption{\label{fig:absorp-1} (Color online) Optical absorbance of
BLG (AB stacked) with applied gating electric field: (a) 0.0 V/nm,
(b) 1.0 V/nm, (c) 2.0 V/nm and (d) 4.0 V/nm. The polarization of
incident light is parallel to the graphene sheet. A 10 meV
Gaussian broadening is applied to all plots. Please pay attention
to the different scale of the above absorbance under different
applied field.}
\end{figure}

First, we will focus on optical absorbance of AB stacked BLG with
the incident light polarized parallel to the graphene sheet. The
calculated optical absorption spectra are presented in
Fig.~\ref{fig:absorp-1}. The optical absorption with a frequency
less than 40 meV is not plotted because intra-band transitions and
Drude factors are important there while we do not include them in
this study. In Fig.~\ref{fig:absorp-1} (a), plotted is the optical
absorbance of AB stacked BLG in the absence of electric field.
There is only one absorption peak around 400 meV that is due to
inter-band transitions between the highest valence band and the
second lowest conduction band and the second highest valence band
and the lowest conduction band, respectively. Besides this optical
absorption peak, the rest part of the absorbance is around 4.7 \%
that is consistent with previous experimental observations
\cite{nair08,heintz08}.

When gating electric field is applied, a new absorption peak shows
up because of the induced finite band gap, as shown in
Fig.~\ref{fig:absorp-1} (b), (c) and (d). By measuring the
position of the sharpest slope of the new absorption peak, we
identify that these values are consistent with our calculated band
gap. Moreover, as applied electric field is stronger, the
lineshape of the first absorption peak changes as well. For
example, in Fig.~\ref{fig:absorp-1} (d), the first absorption peak
is actually a combination of a few peaks. From
Fig.~\ref{fig:band-2} (a), we can understand the origin of the
change of the absorption lineshape is from the Mexican-hat band
structure under strong applied field. These changes of band
dispersions and correspond optical absorption spectra will modify
the effective mass of free carriers and are important to transport
and electro-optical properties of BLG.

\begin{figure}
\includegraphics*[scale=1.05]{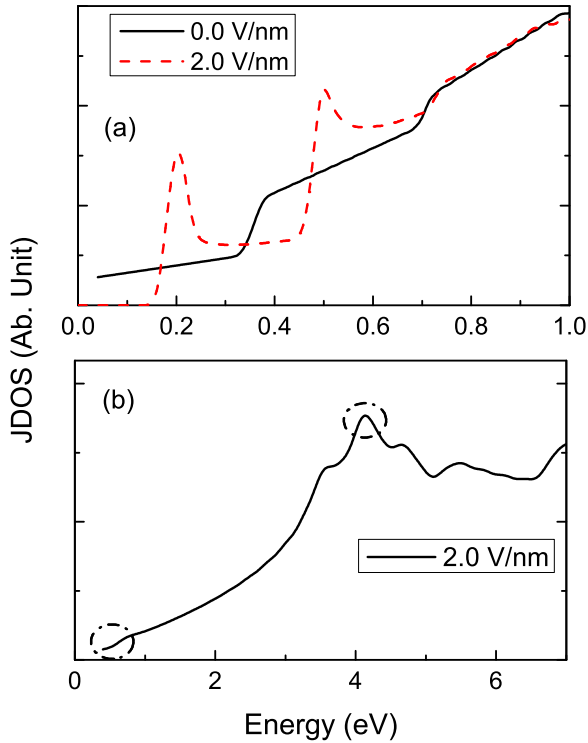}
\caption{\label{fig:JDOS} (Color online) The JDOS of bare and
electrically gated BLG (AB stacked). (a) The JDOS from 0 to 1 eV;
(b) the JDOS from 0 to 7 eV. A 10 meV Gaussian broadening is
applied to (a) and 100 meV Gaussian broadening is applied to (b).}
\end{figure}

It has to be paid attention to that not only the peak position but
also the peak intensity is modified by the applied electric field;
larger field induces a stronger absorption peak. Plotted in
Fig.~\ref{fig:JDOS} is the corresponding joint density of states
(JDOS), which is helpful to understand the modification of the
absorbance. As shown in Fig.~\ref{fig:JDOS} (a), the applied
electric field not only induces a finite band gap but also gives
rise to an enhanced peak at the band edge of the JDOS. This
enhancement of the JDOS increases the absorbance intensity because
more inter-band transitions are available within the certain
frequency regime.

When turning to the second absorption peak, we find a weaker field
dependence, which agrees well with our band structure conclusion
because the second highest valence band and the second lowest
conduction band are not sensitive to the applied field. Therefore,
the shift of this peak is not prominent until the applied electric
field is larger than 2.0 V/nm. Unlike the first absorption peak
whose intensity is significantly enhanced, the intensity of the
second peak does not change much under different applied fields.

\begin{figure}
\includegraphics*[scale=0.85]{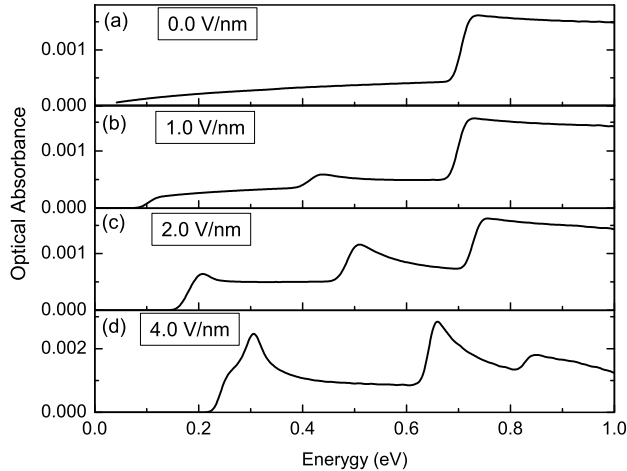}
\caption{\label{fig:absorp-3} (Color online) Optical absorbance of
BLG (AB stacked) with applied gating electric field: (a) 0.0 V/nm,
(b) 1.0 V/nm, (c) 2.0 V/nm and (d) 4.0 V/nm. The polarization of
incident light is perpendicular to the graphene sheet. A 10 meV
Gaussian broadening is applied to all plots. Please pay attention
to the different scale of the above absorbance under different
applied field.}
\end{figure}

In low-dimensional structures, the optical response is strongly
depending on the polarization direction of the incident light. In
BLG, we find the similar phenomenon. In Fig.~\ref{fig:absorp-3},
we present the optical absorbance of the incident light with a
polarization direction perpendicular to the graphene sheet. First,
the optical absorbance in this case is around two orders of
magnitude smaller than that in Fig.~\ref{fig:absorp-1}. This
depolarization effect is interesting and quite different from
those observed in other nanostructures
\cite{catalin04,eric04,yang07}. In those studies, we have to
include the local field factor to obtain the depolarization effect
that is not considered in this study yet.

Second, we see more fine structures from this perpendicular
polarization case. For example, we observe the absorption peaks
originated between the second highest valence band and the second
lowest conduction band in Fig.~\ref{fig:absorp-3} which is not
shown in Fig.~\ref{fig:absorp-1}. Therefore, the energy position
of these two relevant bands can be measured by perpendicularly
polarized optical absorption spectra. Moreover, the intensity of
the first absorption peak shows a stronger dependence on applied
field than that of parallel-polarized cases. For example, its
absorbance increases from 0.03\% to 0.3\% as the field changes
from 0.5 V/nm to 4 V/nm. Therefore, although the magnitude of the
absorbance is much smaller than that of Fig.~\ref{fig:absorp-1},
it provides stronger contrast if advanced experimental techniques
can detect them, which may give better accuracy to identify the
band structure.

\begin{figure}
\includegraphics*[scale=0.85]{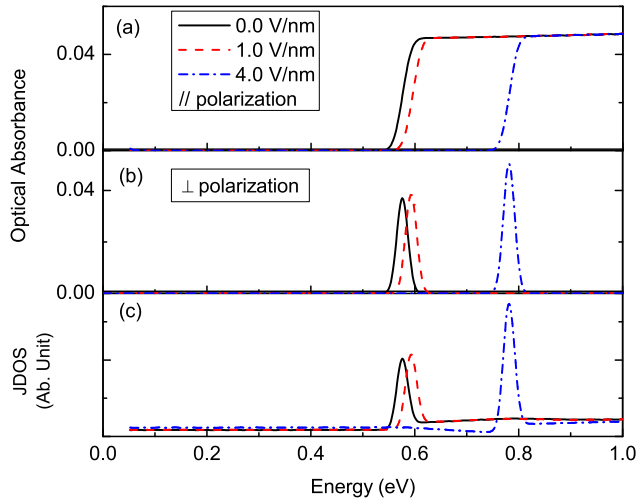}
\caption{\label{fig:absorp-4} (Color online) Optical absorbance
and JDOS of AA stacked BLG under different gating electric field.
(a) Optical absorbance with parallel polarized incident light; (b)
optical absorbance with perpendicularly polarized incident light;
(c) JDOS. A 10 meV Gaussian broadening is applied to all plots.}
\end{figure}

Although AB stacked BLG is theoretically more stable than AA
stacked one, it is interesting to study the optical response of
the later one because the strain, imperfections and grain
boundaries may result in locally AA stacked BLG. Therefore, we
have calculated the optical absorbance of AA stacked BLG and
presented them in Fig.~\ref{fig:absorp-4}. Because of different
symmetries, the AA stacked BLG shows a very different optical
response from that of AB stacked one. In Fig.~\ref{fig:absorp-4}
(a), when the incident light is polarized parallel to the graphene
sheet, the optical absorbance is zero within the frequency range
up to 0.6 eV. This symmetry gap is due to the zero-oscillator
strength between transitions from the highest valence band to the
lowest conduction band \cite{min09}. Beyond that, the optical
absorption is contributed from transitions between the highest
valence band and the second lowest conduction band and the second
highest valence band and the lowest conduction band, respectively.
Interestingly, the optical absorbance above 0.6 eV is nearly a
constant ($\sim$ 4.7\%) that is the same as that of AB stacked
BLG. Therefore, the universal infrared optical conductance of BLG
above 0.6 eV is not sensitive to whether AA or AB stacking style.

Finally, the electric-field effect on the optical response of AA
stacked BLG is weak as shown in Fig.~\ref{fig:absorp-4}. A
significant shift of the absorption edge does not show up until
the applied electric field is larger than 2 V/nm, which is
consistent with our band structure calculations shown in
Fig.~\ref{fig:band-2} (b).

The polarization effect in AA stacked BLG is also quite different
from that of AB stacked one. As shown in Fig.~\ref{fig:absorp-4}
(b), the magnitude of the optical absorbance for the polarization
direction perpendicular to the graphene sheet is comparable to
that with a parallel polarization while a significant
depolarization effect are observed in AB stacked BLG. However,
since the local field effect may depress the perpendicularly
polarized optical absorption spectrum, a significant change of
Fig.~\ref{fig:absorp-4} (b) may happen after including
many-electron effects. Interestingly, when we compare the optical
absorbance with the corresponding JDOS that is shown in
Fig.~\ref{fig:absorp-4} (c), we find the perpendicularly polarized
optical absorbance has a similar peak structure around 0.6 eV as
that of the JDOS. This fact suggests that perpendicularly
polarized absorbance is a better choice to measure the JDOS in AA
stacked BLG.

\section{Excitonic Effects on Infrared Optical Absorption Spectra}

It is known that many-electron effects, like electron-hole
interactions \cite{hanke,rohlfing}, are of importance in
determining the optical response of low-dimensional carbon
materials \cite{catalin-1,yangnano,yang08,prezzi08}. In
particular, a previous first-principles calculation with
many-electron effects included has revealed enhanced electron-hole
interactions in the optical absorption spectrum of BLG around 5 eV
\cite{yang09}. Therefore, questions to our DFT-calculated infrared
optical absorption spectra are if many-electron effects will play
an important role there and if they will qualitatively change our
result for BLG with a finite band gap.

Usually, there are two important factors to dictate excitonic
effects on the optical response of solids. One is the screening
between electrons and holes; a smaller band gap means a stronger
screening and weaker electron-hole interactions. The other one is
the number of available electron-hole pair configurations within a
certain energy regime. Within the Tamm-Dancoff approximation
\cite{tamm}, an exciton state can be written as
\begin{equation}\label{2}
\chi_S (x_e,x_h) = \sum_k \sum_v^{hole} \sum_c^{elec} A_{vck}^S
\psi_{c,k}(x_e) \psi^{\ast}_{v,k}(x_h),
\end{equation}
where $A_{vck}$ is the exciton amplitude, $\psi_{c,k}(x_e)$ and
$\psi_{v,k}(x_h)$ are the electron and hole states, respectively.
From this formula, it is easy to see that a stronger bound exciton
state needs more electron-hole pair configurations to form a
localized state. Therefore, flatter bands are preferred to form
enhanced excitonic effects, which is consistent with the
hydrogenic model because larger effective mass gives rise to a
stronger binding energy of excitons accordingly.

For electrically gated BLG, an important consideration is the
small number of available electron-hole pair states within the
infrared frequency regime because of the sharp slope of band
dispersion close to the Dirac point of BLG. In Fig.~\ref{fig:JDOS}
(b), we have marked the JDOS around the band gap and the peak
around 4 to 5 eV. Since the JDOS around the induced band gap is
much smaller than that around 4 to 5 eV, we expect excitonic
effects around the band gap is much smaller than those around 4 to
5 eV. However, since the induced band gap is relatively small,
excitons with a small binding energy (a few tenths of an eV) can
modify optical spectra \cite{ch09} although they will not
significantly change main conclusions of our calculation. To
justify this open question about the excitonic effects in
electrically gated BLG, more accurate first-principles studies are
necessary, which are beyond this paper. In addition, we suggest
future experiments to be performed by measuring the lineshape of
absorption peaks as what had been done in metallic Carbon
nanotubes \cite{jack08} to check the existence of bound excitons
in electrically gated BLG.

\section{Summary}

In conclusion, we have performed first-principles calculations on
the electronic structure and optical absorption spectra of
electrically gated BLG. The electric-field dependence of band gaps
is evaluated. Our calculated result is partially in good agreement
with recent experiments. We believe self-energy corrections are
important although experimental substrate effects can depress it
when the applied field is weak.

The optical absorbance is calculated within the single-particle
transition picture. Absorption peaks, lineshapes and intensity are
found to be strongly depending on the applied electric field. The
polarization effect of the incident light and stacking styles of
BLG are studied as well, which provide efficient ways to detect
the atomic and electronic structure of BLG. Finally, excitonic
effects are discussed and possible experiments are suggested to
verify our calculations.

\section*{ACKNOWLEDGMENTS}

We thank Cheol-Hwan Park and Anders E. Carlsson for fruitful
discussions. Computational resources are provided by Lonestar of
Teragrid at the Texas Advanced Computing Center .


\end{document}